\documentclass[10pt,aps,prl,twocolumn,nopacs,amsmath,amssymb,superscriptaddress]{revtex4}

\usepackage[ansinew]{inputenc}
\usepackage[type1]{libertine}                                        
\usepackage{textcomp}
\usepackage[scaled=.85]{beramono}
\usepackage[libertine,cmintegrals,cmbraces,vvarbb,slantedGreek]{newtxmath}
\usepackage[scr=boondoxo]{mathalfa}
\usepackage{bm}
\usepackage[lf]{carlito}

\usepackage{braket}

\usepackage{bm} 
\usepackage{graphicx}

\makeatletter
\def\NAT@bibsetnum#1{
 \setlength{\topsep}{\z@}
 \NATx@bibsetnum{#1}
}
\renewenvironment{thebibliography}[1]{
 \NAT@thebibliography{#1}
 \@clubpenalty\clubpenalty
 \let\@TBN@opr\present@bibnote
 \@FMN@list
}{
 \edef\@currentlabel{\arabic{NAT@ctr}}
 \NAT@endthebibliography
 \global\let\auto@bib\@empty
}
\newcommand*{\supplementarystart}{
  \let\@FMN@list\relax 
  \close@column@grid
  \clearpage
  \onecolumngrid
  \setcounter{enumiv}{0} 
  \setcounter{equation}{0} 
  \setcounter{figure}{0} 
  \setcounter{table}{0} 
  \setcounter{page}{1} 
  \c@secnumdepth=4
  \renewcommand{\theequation}{S\arabic{equation}} 
  \renewcommand{\bibnumfmt}[1]{[S##1]} 
  \renewcommand{\citenumfont}[1]{S##1} 
  \renewcommand{\thefigure}{S\arabic{figure}} 
  \renewcommand{\thetable}{S\Roman{table}} 
  \renewcommand{\thepage}{S\arabic{page}} 
}
\makeatother

\begin{document}

\title{Band dispersion of graphene with structural defects}

\author{Piotr Kot}
\affiliation{Max-Planck-Institut f\"ur Festk\"orperforschung, 70569 Stuttgart, Germany}

\author{Jonathan Parnell}
\affiliation{University of British Columbia, Vancouver, Canada}

\author{Sina Habibian}
\affiliation{University of British Columbia, Vancouver, Canada}

\author{Carola Straßer}
\affiliation{Max-Planck-Institut f\"ur Festk\"orperforschung, 70569 Stuttgart, Germany}

\author{Pavel M.\ Ostrovsky}
\affiliation{Max-Planck-Institut f\"ur Festk\"orperforschung, 70569 Stuttgart, Germany}
\affiliation{L.\ D.\ Landau Institute for Theoretical Physics RAS, 119334 Moscow, Russia}

\author{Christian R.\ Ast}
\email[Corresponding author; e-mail:\ ]{c.ast@fkf.mpg.de.}
\affiliation{Max-Planck-Institut f\"ur Festk\"orperforschung, 70569 Stuttgart, Germany}

\begin{abstract}
We study the band dispersion of graphene with randomly distributed structural defects using two complementary methods, exact diagonalization of the tight-binding Hamiltonian and implementing a self-consistent $T$ matrix approximation. We identify three distinct types of impurities resulting in qualitatively different spectra in the vicinity of the Dirac point. First, resonant impurities, such as vacancies or 585 defects, lead to stretching of the spectrum at the Dirac point with a finite density of localized states. This type of spectrum has been observed in epitaxial graphene by photoemission spectroscopy and discussed extensively in the literature. Second, nonresonant (weak) impurities, such as paired vacancies or Stone-Wales defects, do not stretch the spectrum but provide a line broadening that increases with energy. Finally, disorder that breaks sublattice symmetry, such as vacancies placed in only one sublattice, open a gap around the Dirac point and create an impurity band in the middle of this gap. We find good agreement between the results of the two methods and also with the experimentally measured spectra.
\end{abstract}

\maketitle

Graphene presents high potential for providing the next generation of electronic materials due to its strictly two-dimensional character as well as its high electron mobility. It has demonstrated high design flexibility, such as doping by atoms or molecules, efficient decoupling from an underlying substrate, or high tensile strength for flexible electronics\,\cite{graphene_doped, graphene_decouple, graphene_flexible}. Several theoretical proposals as well as experiments are concerned with enhancing the spin-orbit coupling to open a band gap or inducing spin-splitting\,\cite{graphene_so1, graphene_so2}. Even a possible transition to a superconducting state has been proposed\,\cite{graphene_supercond}. However, one of the most important prerequisites for graphene to become a base material for future electronics concerns the opening of a band gap, which has not been successfully demonstrated so far. The linear crossing of the bands near the Dirac point is protected by symmetry, because the two sublattices are equivalent. In order to open a band gap, this sublattice symmetry has to be broken.

Angle-resolved photoemission spectroscopy (ARPES) is the most direct method to probe the electronic structure experimentally. Numerous studies have examined the band structure of graphene near the Dirac point\,\cite{coletti_charge_2010, bostwick_experimental_2009, bostwick_observation_2010, gierz2008atomic, gierz2011illuminating, gierz2012graphene, graphene_dirac1, graphene_dirac2, graphene_dirac3, bostwick2007quasiparticle}. Several of these studies observe an elongated region near Dirac point as if the two touching cones are pulled apart, stretched but without tearing apart. Such occurrences have been discussed extensively in literature and depending on the specific environment of the graphene, were attributed to imperfections in the graphene, interactions with the substrate, or the opening of a band gap\,\cite{graphene_gap, graphene_gap2, walter2011effective, bostwick2007quasiparticle, graphenecoll, zhou_departure_2008}. One observation common to all instances of the stretched Dirac point is the residual spectral weight that is still present at lowest energies. It needs to be understood in more detail in order to judge, if and under what conditions it can be referred to as an actual gap.

In this letter, we present a real-space tight-binding calculation modeling different kinds of structural defects randomly distributed over a graphene sample. We show that in all instances, except the case of vacancies placed in a single sublattice, there is no band gap opening near the Dirac point. The spectrum near the Dirac point is either almost unchanged or exhibits a stretched Dirac point with broadened states, which resemble experimentally observed band dispersion. Complementing our tight-binding model with a self-consistent $T$ matrix approximation (SCTMA) calculation\,\cite{graphene_disorder, supplementary}, we show that point defects in graphene are either a resonant or nonresonant type\,\cite{graphene_resonant, graphene_resonant2}. It is resonant defects that produce a dispersion with broadened states near the Dirac point resembling the results of the tight-binding calculation as well as experimental findings. Due to the remarkable similarity between the results of the SCTMA, numerical simulations of the tight-binding model, and the experiment, we conclude that the broadened states measured in epitaxial graphene must at least in part be caused by specific resonant defects in graphene. On the other hand, nonresonant defects are shown to weakly affect the dispersion near the Dirac point, which is also confirmed by the tight-binding model calculation. These findings lead to the conclusion that it is close to impossible to open a band gap in graphene by virtue of defects only.

We employed a second nearest-neighbor real-space tight-binding model with the following Hamiltonian:
\begin{equation}
 H
  = -\sum_{\langle ij\rangle} |i\rangle t \langle j| + \sum_{\langle\!\langle ij\rangle\!\rangle} |i\rangle t' \langle j|.
\end{equation}
Here the indices $i$ and $j$ label individual atoms with one $p_z$-orbital per atom. The parameters $t$ and $t'$ are hopping amplitudes between first and second nearest neighbors, respectively. The sums with single and double angular brackets run over first and second nearest neighbors, respectively. We have typically built the Hamiltonian for a rectangular supercell of 160,000 carbon atoms with periodic boundary conditions and add randomly distributed structural defects with a given concentration. The Hamiltonian is then exactly diagonalized numerically and the resulting real-space wave functions are converted to momentum-space. This yields the complete set of eigenenergies $E_n$ and corresponding eigenfunctions $\psi_n(\mathbf{p})$. The spectral weight function is then computed as
\begin{equation}
 A(E, \mathbf{p})
  = \sum_n |\psi_n(\mathbf{p})|^2 \delta(E - E_n).
\end{equation}

The code was programmed in Matlab using the built-in routines for calculating eigenvalues and eigenvectors as well as Fourier transforms. The defects were introduced by suppressing the hopping between particular randomly chosen lattice sites. The atom positions near the defects have not been relaxed, except for the Stone-Wales (SW) defect, which involves repositioning of two carbon atoms\,\cite{graphene_defects}. The concentration of defects, $n_\text{imp}$, is the ratio of the number of carbon atoms taken out or displaced to the total number of carbon atoms in the lattice. The large number of lattice sites used in the calculation is necessary for statistical reasons to increase the number of eigenstates in the region of low density of states (DOS) near the Dirac point as well as to have a large enough number of randomly distributed defects in the supercell. For the calculations presented in this letter the values of the parameters are $t=3.033$\,eV and $t'=0.2$\,eV\,\cite{graphene_transport}.

In the alternative SCTMA calculation, we start with the exact Green's function on a honeycomb lattice at zero energy. Only nearest-neighbor hopping terms are taken into account with $t=3.033$\,eV. Any individual structural defect considered in our calculation perturbs at most six neighboring sites of the lattice, hence we describe it with an exact $6 \times 6$ $T$ matrix. We then convert this exact zero-energy $T$ matrix to the basis of $4$-component spinors governed by the continuous low-energy massless Dirac Hamiltonian with two valleys\,\cite{graphene_Diracham}. The converted $T$ matrix is averaged over all possible positions and orientations of the defect and a non-zero energy is introduced as a perturbation. The averaged Green's functions of graphene with a finite concentration of defects, $n_\text{imp}$, acquire the self-energy $\Sigma = (n_\text{imp}/A) \langle T(E - \Sigma) \rangle$. Here we have also included the same self energy in the argument of the impurity $T$ matrix thus introducing the self-consistency equation. The spectral weight is calculated from the self energy as
\begin{equation}
 A(E, \mathbf{p})
  = -\frac{2}{\pi} \mathop{\mathrm{Im}} \left[
      \frac{1}{E - \Sigma - v p} + \frac{1}{E - \Sigma + v p}
    \right]
\end{equation}
and is compared to the results of the tight-binding model and to the experimentally measured dispersion. Further details of the SCTMA calculation can be found in the Supplemental Material\,\cite{supplementary}.

\begin{figure}
\centerline{ \includegraphics[width = \columnwidth]{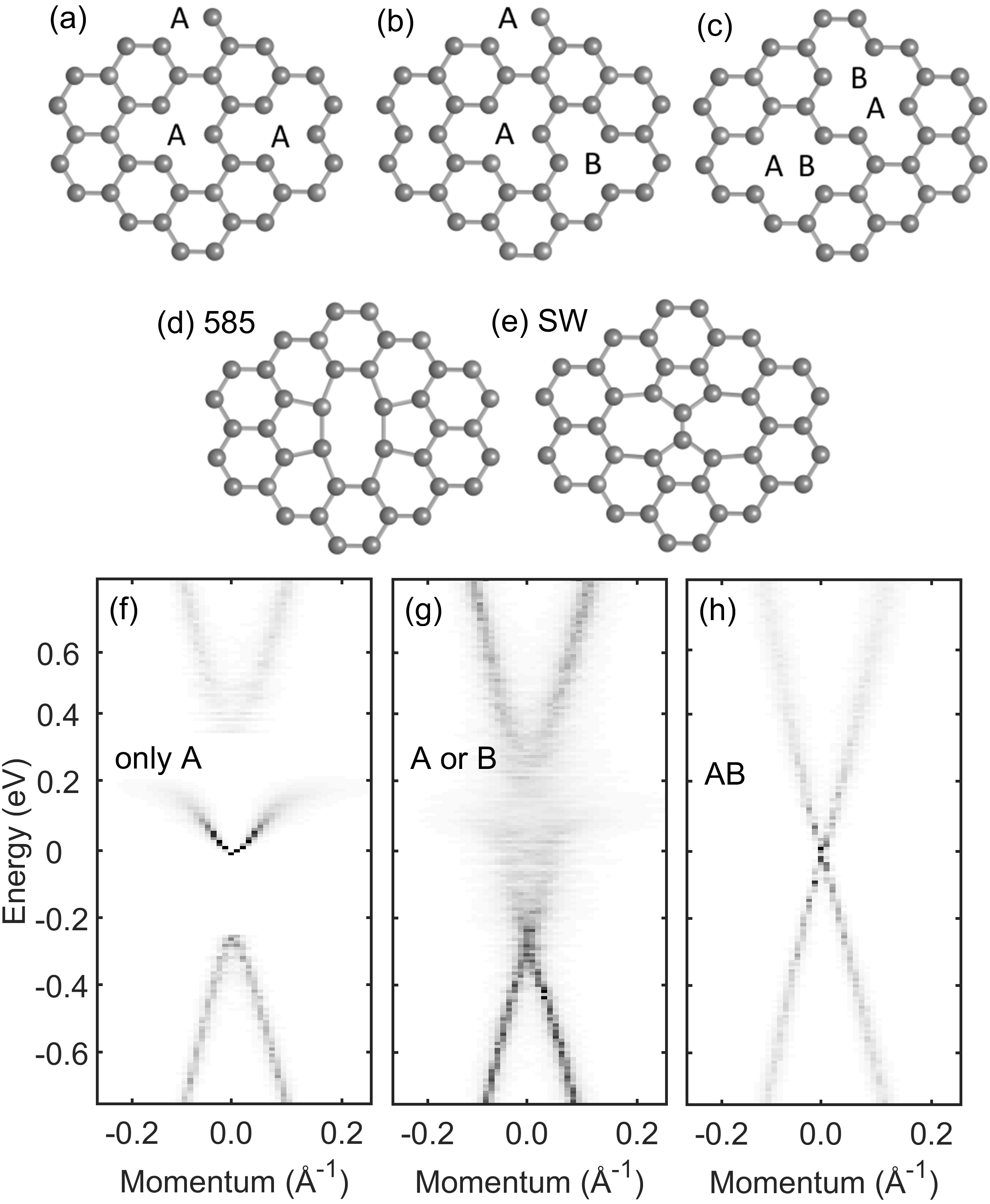}}
\caption{Different types of structural defects: (a) vacancies in a single sublattice, (b) vacancies in both sublattices, (c) double vacancies, (d) 585 defect with two sites removed and two bonds reconstructed, (e) Stone-Wales defect with two adjacent atoms rotated by 90$^{\circ}$\,\cite{graphene_defects}. Electron spectrum computed from exact diagonalization of the tight-binding model: (f) for vacancies in a single sublattice, (g) for vacancies in both sublattices, (h) for double vacancies.}
\label{fig:Def_Overview_AB}
\end{figure}

We consider five different types of structural defects illustrated in Fig.\,\ref{fig:Def_Overview_AB}. They show the effect of sublattice symmetry breaking and the qualitative difference between resonant and nonresonant defects. The first three types of impurities are vacancies that are either distributed in a single (A) sublattice, equally in both sublattices (A and B), or paired (whole AB unit cells removed). These defects are shown in Fig.\,\ref{fig:Def_Overview_AB}(a--c). While vacancies are not feasible in graphene, they provide a good model for adatoms attached to individual lattice sites inducing a strong on-site potential\,\cite{graphene_resonant, Wehling07}. Vacancies allow us to probe sublattice symmetry breaking and to demonstrate its effect on the band dispersion\,\cite{graphene_resonant2, graphene_Diracham, Wehling07, graphene_vacancies_numerics, graphene_vacancies_theory}. We also consider two other structural point defects, that have been experimentally observed in epitaxial graphene\,\cite{lauffer2008atomic} and are shown in Fig.\,\ref{fig:Def_Overview_AB}(d, e). The 585 defect, Fig.\,\ref{fig:Def_Overview_AB}(d), is similar to the AB paired vacancy but with two reconstructed bonds. The SW defect is shown in Fig.\,\ref{fig:Def_Overview_AB}(e) and involves a 90$^{\circ}$ rotation of a bond between two adjacent atoms, along with the rearranging of the hopping terms around them.

Numerically computed spectra for distributions of vacancies are shown in Fig.\,\ref{fig:Def_Overview_AB}(f--h). We see that vacancies placed in only one sublattice, Fig.\,\ref{fig:Def_Overview_AB}(f), open a gap in the Dirac spectrum with an additional midgap band\,\cite{graphene_vacancies_theory}. We will discuss the origin of this extra band below. Removing atoms randomly from both sublattices, as shown in Fig.\,\ref{fig:Def_Overview_AB}(g), induces additional spectral weight in the vicinity of the Dirac point with momentum broadening that gets stronger closer to zero energy. At higher energies we see a band structure resembling ``elongated'' Dirac points. Finally, removing adjacent atoms, Fig.\,\ref{fig:Def_Overview_AB}(h), does not noticeably affect the spectrum apart from an energy dependent broadening of the line width.  Among these three qualitatively different electronic spectra, only the first one (vacancies placed in one sublattice, Fig.\,\ref{fig:Def_Overview_AB}(a)) provides a true band gap, Fig.\,\ref{fig:Def_Overview_AB}(f).

\begin{figure}
\centerline{\includegraphics[width = 1.0\columnwidth]{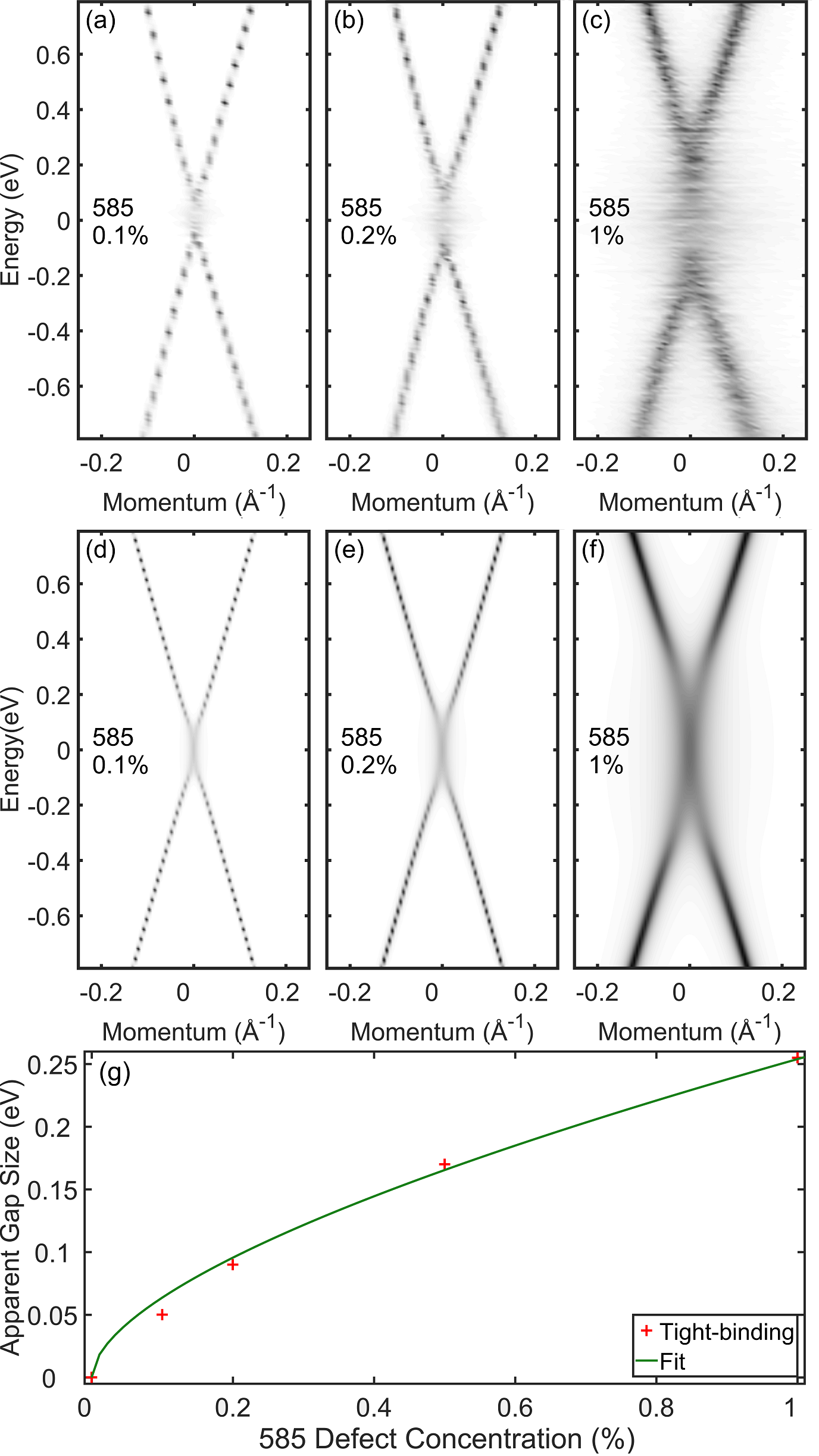}}
\caption{Band structure of graphene with 585 defects obtained from (a--c) exact diagonalization of the tight-binding Hamiltonian and (d--f) SCTMA calculation. A significant ``stretching'' near the Dirac point is already visible for $0.1\%$ defect density. Such spectrum is typical for resonant impurities. (g) An amount of the Dirac point stretching (in energy) as a function of defect concentration compared to Eq.\,(\protect\ref{eq:gap}) (shown with solid line).}
\label{fig:Defect585}
\end{figure}

The structural defects shown in Fig.\,\ref{fig:Def_Overview_AB}(d, e) also demonstrate qualitatively different spectra similar to Fig.\,\ref{fig:Def_Overview_AB}(g, h), respectively. We analyze them analytically in the framework of the SCTMA; see Supplemental Material\,\cite{supplementary}. For the 585 defect we find the following equation for the self energy:
\begin{equation}\label{eq:res}
 \Sigma
  = \frac{-\beta n_\text{imp}}{(E - \Sigma) \log(-i(E - \Sigma)/t)}.
\end{equation}
This form is the result of a divergent zero-energy $T$ matrix\,\cite{graphene_disorder, supplementary}. Such defects are known as resonant. For the case of the 585 defect $\beta=12.5$\,eV$^2$. In Fig.\,\ref{fig:Defect585} we compare the spectral weight obtained from direct numerical diagonalization of the disordered lattice model (panels (a--c)) with the solution of the self-consistency Eq\,(\ref{eq:res}) (panels (d--f)). For both calculations, the resulting structure near the Dirac point is very similar to experimentally measured graphene dispersion. Stretching of the spectrum near the Dirac point can be estimated from Eq.\,(\ref{eq:res}) as
\begin{equation}\label{eq:gap}
 \Delta
  = \sqrt{\frac{2 \beta n_\text{imp}}{|\ln (c n_\text{imp})|}},
\end{equation}
where $c$ is a fitting parameter. We plot the size of the smeared region around the Dirac point in Fig.\,\ref{fig:Defect585}(g) along with the fit ($c = 2.0712 \pm 0.38$).

\begin{figure}
\centerline{\includegraphics[width = \columnwidth]{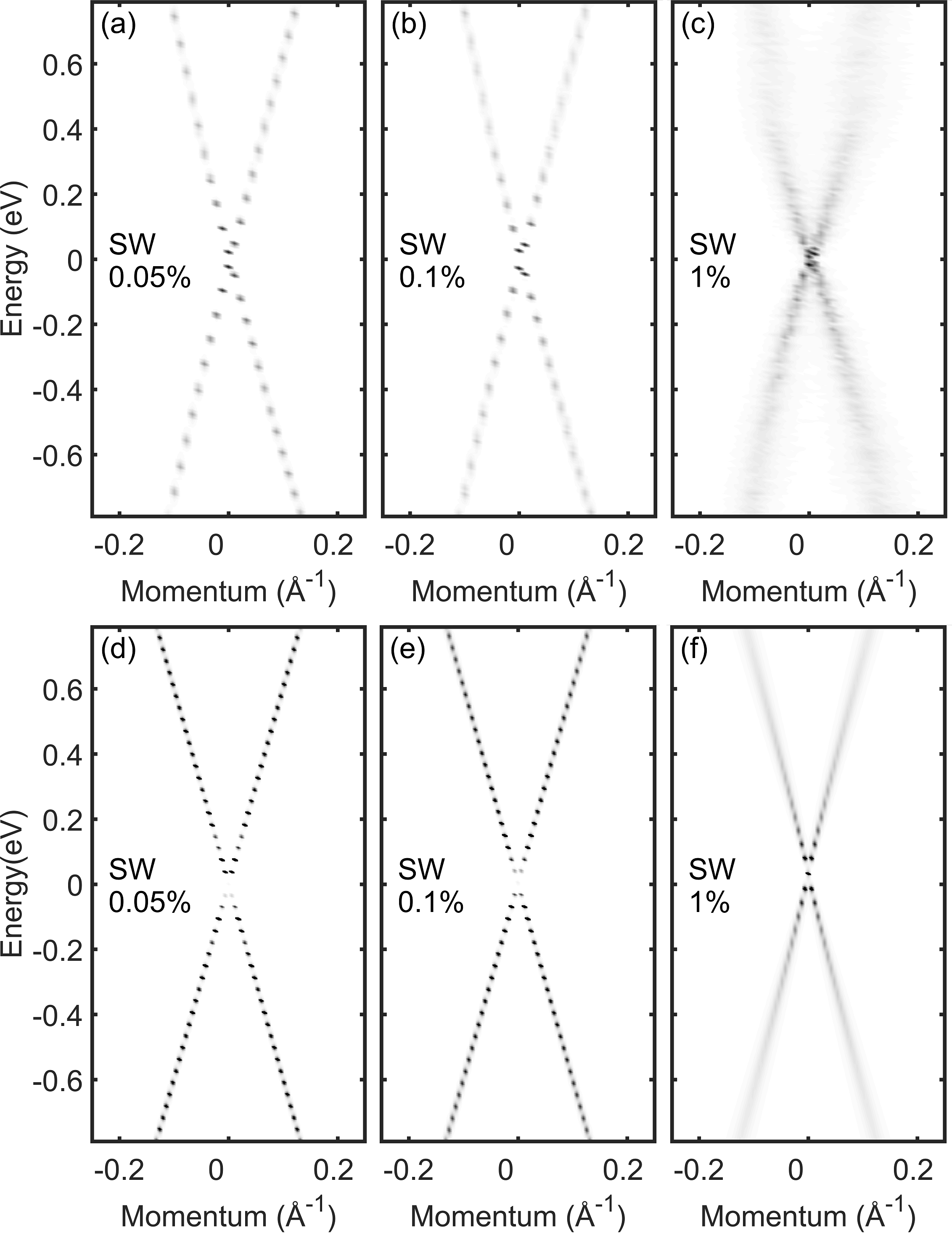}}
\caption{
Electron spectrum of graphene with SW defects obtained from (a--c) exact diagonalization of the tight-binding Hamiltonian and (d--f) SCTMA calculation. In contrast to the 585 defect, cf. Fig.\,\ref{fig:Defect585}, stretching near the Dirac point does not occur. Instead the line width broadening gradually increases away from the Dirac point. This behavior is typical for nonresonant impurities.}
\label{fig:DefectSW}
\end{figure}

Another type of structural defects (SW, Fig.\ \ref{fig:Def_Overview_AB}(e)) belongs to the nonresonant case. In this case, the SCTMA provides the following self energy equation\,\cite{supplementary}:
\begin{equation}\label{eq:nores}
 \Sigma
  = n_\text{imp} [\alpha (E - \Sigma) \log(-i(E - \Sigma)/t)].
\end{equation}
The nonresonant case is typically what is found for most point defects and for the SW defect $\alpha=6.85$. The band dispersion from the direct numerical diagonalization and from the SCTMA are shown in Fig.\,\ref{fig:DefectSW}. There is no apparent ``elongated'' Dirac point even at relatively high concentrations of impurities. Instead an energy-dependent line broadening gets stronger away from the Dirac point. This is in contrast to the relatively uniform broadening found in the band dispersion with resonant impurities. We see that both for resonant 585 defects and non-resonant SW defects, the results of direct diagonalization and SCTMA calculation show a remarkable agreement.

The two forms of the self-energy, Eqs.\,(\ref{eq:res}) and (\ref{eq:nores}), are the only two possibilities for point defects (as long as sublattice symmetry is preserved). In the resonant case, the zero-energy $T$ matrix diverges hence it can be approximated as $T \sim 1/E$ (up to a logarithmic factor). This leads to the self-consistency equation of the form in Eq\,(\ref{eq:res}). In the nonresonant case, small-energy expansion of the $T$ matrix starts with a nonessential constant (it can be absorbed in the chemical potential) and a linear term leading to Eq.\ (\ref{eq:nores}). Both resonant and nonresonant cases are in a good agreement with our direct numerical simulations of the tight-binding model. When the sublattice symmetry is broken, self energy is not a number anymore but rather an operator in the sublattice space. Then more possibilities beyond Eqs.\,(\ref{eq:res}) and (\ref{eq:nores}) emerge\,\cite{graphene_resonant2}. Vacancies distributed in only one sublattice, Fig.\,\ref{fig:Def_Overview_AB}(a, f), is one possible illustration of this effect.

\begin{figure}
\centerline{\includegraphics[width = \columnwidth]{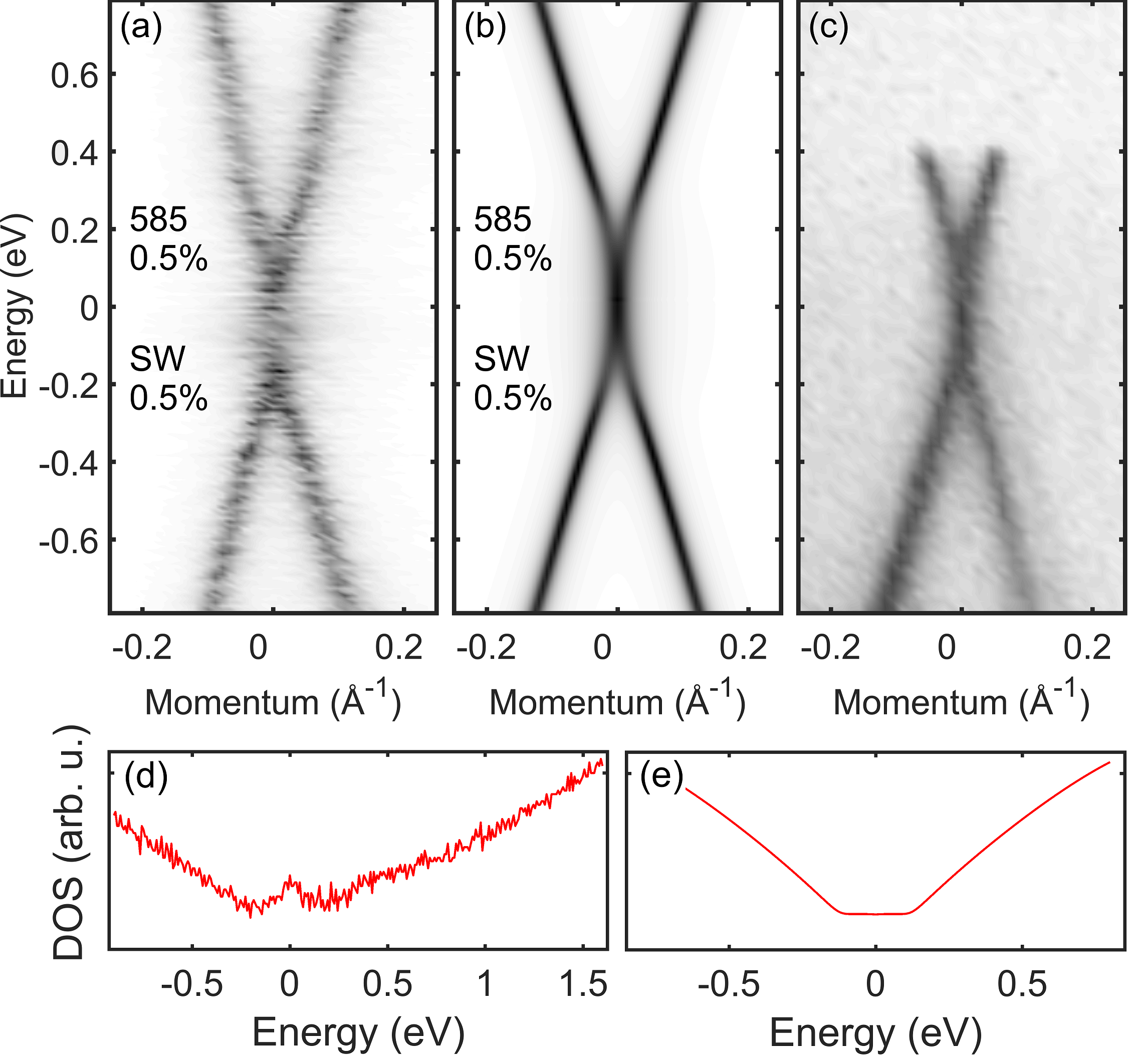}}
\caption{Electron spectrum of graphene with equal concentration of 585 and SW defects from (a) exact diagonalization of the tight-binding Hamiltonian and (b) SCTMA calculation. Vicinity of the Dirac point is dominated by the resonant 585 impurities with the characteristic stretching of the spectrum, cf.\ Fig.\,\protect\ref{fig:Defect585}. Line width away from the Dirac point is broadened mainly by the weak SW defects as in Fig.\,\protect\ref{fig:DefectSW}. (c) ARPES spectrum of epitaxial graphene\,\cite{graphene_gap, graphene_gap2, walter2011effective, bostwick2007quasiparticle, graphenecoll, zhou_departure_2008}. (d, e) Average DOS corresponding to the spectra in (a, b).}
\label{fig:Defect585SW}
\end{figure}

We also consider a mixture of 585 and SW defects in a single sample to gain insight on how these defects interact and to have an approximation of realistically disordered graphene. In Fig.\,\ref{fig:Defect585SW}, we compare the band structure for an equal amount of 585 and SW defects obtained from direct diagonalization (panel (a)) and the SCTMA calculation (panel (b)) with the ARPES data of epitaxial graphene (panel (c))\,\cite{graphene_gap, graphene_gap2, walter2011effective, bostwick2007quasiparticle, graphenecoll, zhou_departure_2008}. The average DOS is shown in Fig.\,\ref{fig:Defect585SW}(d, e) for the tight-binding model and SCTMA, respectively. In the tight-binding model we implement an equal number of 585 and SW defects in the lattice while for the SCTMA calculation the self-consistency equation contains the sum of the right-hand sides of Eqs.\,(\ref{eq:res}) and (\ref{eq:nores}). The DOS is a momentum integral of the previously calculated spectral weight.

The results in Fig.\,\ref{fig:Defect585SW}(a, b) not only show a striking resemblance to each other but also match the ``elongation'' found in experimentally measured epitaxial graphene dispersions as seen in Fig.\,\ref{fig:Defect585SW}(c). The DOS shows a nearly constant region near the Dirac point while further away it matches the linear dispersion of states expected for graphene. This finite DOS near the Dirac point is dominated by the effect from resonant defects whose $T$ matrix diverges at zero energy. The DOSs also show an excellent agreement between the SCTMA and the direct tight-binding model.

From our calculations, we can make several conclusions and comment on some new insight. Firstly, the only possibility for a gap in the spectrum is when the sublattice symmetry is broken, for example in the case of vacancies in a single sublattice, Fig.\,\ref{fig:Def_Overview_AB}(f). Mathematically, this can be understood by representing the tight-binding Hamiltonian in the matrix form as \cite{graphene_vacancies_theory}
\begin{equation}\label{eq:H}
 H
  = \begin{pmatrix}
      0 & h \\
      h^{\dagger} & 0 \\
    \end{pmatrix}.
\end{equation}
Here, the block $h$ contains hopping terms from one sublattice to the other and $h^{\dagger}$ defines the opposite hopping. Second nearest neighbor hopping is temporarily disregarded. Removing atoms from only one sublattice causes $h$ and $h^{\dagger}$ to be non-square blocks. The matrix (\ref{eq:H}) has a number of zero eigenvalues equal (or larger\,\cite{Evers16}) to the difference of the number of A and B sites. Hence the DOS exhibits a delta-peak at zero energy with a gap opening around it. This gap is a result of statistical repulsion between zero and non-zero eigenstates of the random Hamiltonian matrix\,\cite{graphene_vacancies_numerics, graphene_vacancies_theory}. When the second nearest neighbor hopping is taken into account, the zero-energy eigenstates rearrange into a dispersive midgap band found in Fig.\,\ref{fig:Def_Overview_AB}(f). These results suggest that selective removal of (or chemical bonding to) carbon atoms from a given sublattice may be the only way to realize gapped graphene experimentally.

The second insight is about the origin of ``elongated'' Dirac point that has been measured experimentally in epitaxial graphene\,\cite{graphene_gap, graphene_gap2, walter2011effective, bostwick2007quasiparticle, graphenecoll, zhou_departure_2008}. The SCTMA shows that a point defect in graphene can either be resonant or nonresonant, where the geometry of a defect affects only the parameters $\alpha$ or $\beta$. Due to the remarkable similarity between experimentally measured ``elongated'' Dirac point (Fig.\,\ref{fig:Defect585SW}(c)), the tight-binding result (Fig.\,\ref{fig:Defect585SW}(a)) and the SCTMA result (Fig.\,\ref{fig:Defect585SW}(b)), we conclude that this ``elongation'' is caused, at least in part, by resonant defects. We have shown that 585 defects are resonant and provide an apparent stretching of the Dirac point. At the same time, it is known that 585 defects are common in epitaxial graphene\,\cite{lauffer2008atomic}. We thus conclude that the ``elongation'' of the Dirac point observed in many graphene samples is the result of resonant defects and 585 defects in particular. Regarding the prospective band gap in epitaxial graphene, our calculations explicitly show that there are states in the Dirac point region, leading us to conclude that the ``elongation'' of the Dirac point can not be considered as a gap. Furthermore, any concentration of resonant defects will increase the number of states at low energies. Nevertheless, the stretching of the spectrum near the Dirac point creates an energy range were electrons are localized in real space. This phenomenon may be used to open a mobility gap around the Dirac point in graphene and realize an insulating state\,\cite{graphene_localization}.

In summary, we implemented a simple real-space tight-binding model that allowed us to calculate the band dispersion of graphene with defects. We found that a band gap can only be induced when the sublattice symmetry is broken as shown in Fig.\,\ref{fig:Def_Overview_AB}(f). Looking at more realistic defects, we found that the 585 defect creates an ``elongated'' Dirac point (Fig.\,\ref{fig:Defect585}) similar to those found in experimentally measured spectra of epitaxial graphene. Graphene with SW defects has a qualitatively different spectrum without apparent stretching (Fig.\,\ref{fig:DefectSW}). We conclude that the experimentally observed ``elongated'' Dirac point is due to the 585 defects and the length of stretching scales as the square root of their concentration. We have also developed a SCTMA theory\,\cite{supplementary} which allowed us to classify point defects as either resonant or nonresonant and showed a remarkable agreement with the direct lattice calculations.  The SCTMA model provides further insights into the nature of the broadened states measured in epitaxial graphene, showing that the ``elongated'' region can not be considered a spectral gap. At the same time, disorder can lead to localization of the states near the Dirac point and hence to a mobility gap\,\cite{graphene_localization}.

\supplementarystart

\centerline{\bfseries\large ONLINE SUPPLEMENTAL MATERIAL}
\vspace{6pt}
\centerline{\bfseries\large Electron spectrum of graphene with structural defects}
\vspace{6pt}
\centerline{Piotr Kot, Jonathan Parnell, Sina Habibian, Carola Straßer, Pavel M.\ Ostrovsky, and Christian R.\ Ast}

\begin{quote}
In this Supplemental Material we provide a detailed derivation of the self-consistent $T$ matrix approximation for structural defects in graphene. We derive exact analytical zero-energy $T$ matrices for Stone-Wales and 585 defects within the tight-binding model. These results are then translated to the effective low-energy description of graphene with the Dirac Hamiltonian. Finally, averaging with respect to positions and orientations of the defects is performed within the self-consistent approximation.
\end{quote}

\subsection{Tight-binding model}

We describe electrons in graphene within the nearest neighbor tight-binding model (second nearest neighbor hopping, cf.\ Eq.\ (1) of the main text, will be neglected throughout this calculation). Wave functions $\Psi(\mathbf{r})$ are defined on the honeycomb lattice with the lattice spacing $a = 2.46\,\text{\AA}$; the spatial argument $\mathbf{r}$ takes the corresponding discrete values. We assume the hopping amplitude of $t = 3.033\,\text{eV}$ as in the main text of the paper. The tight-binding Hamiltonian acts according to
\begin{equation}
 \hat h \Psi(\mathbf{r})
  = -t \sum_{s=0,1,2} \Psi(\mathbf{r} + \zeta_\mathbf{r} \bm{\delta}_s).
\end{equation}
Here $\mathbf{r}$ refers to lattice sites, $\zeta_\mathbf{r} = \pm 1$ is a sign function distinguishing A and B sublattices, and vectors $\bm{\delta}_s$ refer to the three nearest neighbors
\begin{equation}
 \bm{\delta}_s
  = \frac{a}{\sqrt{3}} \begin{pmatrix} \cos(\alpha + 2\pi s/3) \\ \sin(\alpha + 2\pi s/3) \end{pmatrix}.
\end{equation}
The angle $\alpha$ defines orientation of the crystal with respect to coordinate axes. The two vectors of the elementary lattice translations are $\bm{\delta}_1 - \bm{\delta}_0$ and $\bm{\delta}_2 - \bm{\delta}_0$.

\begin{figure}
\centerline{\includegraphics[width=0.7\textwidth]{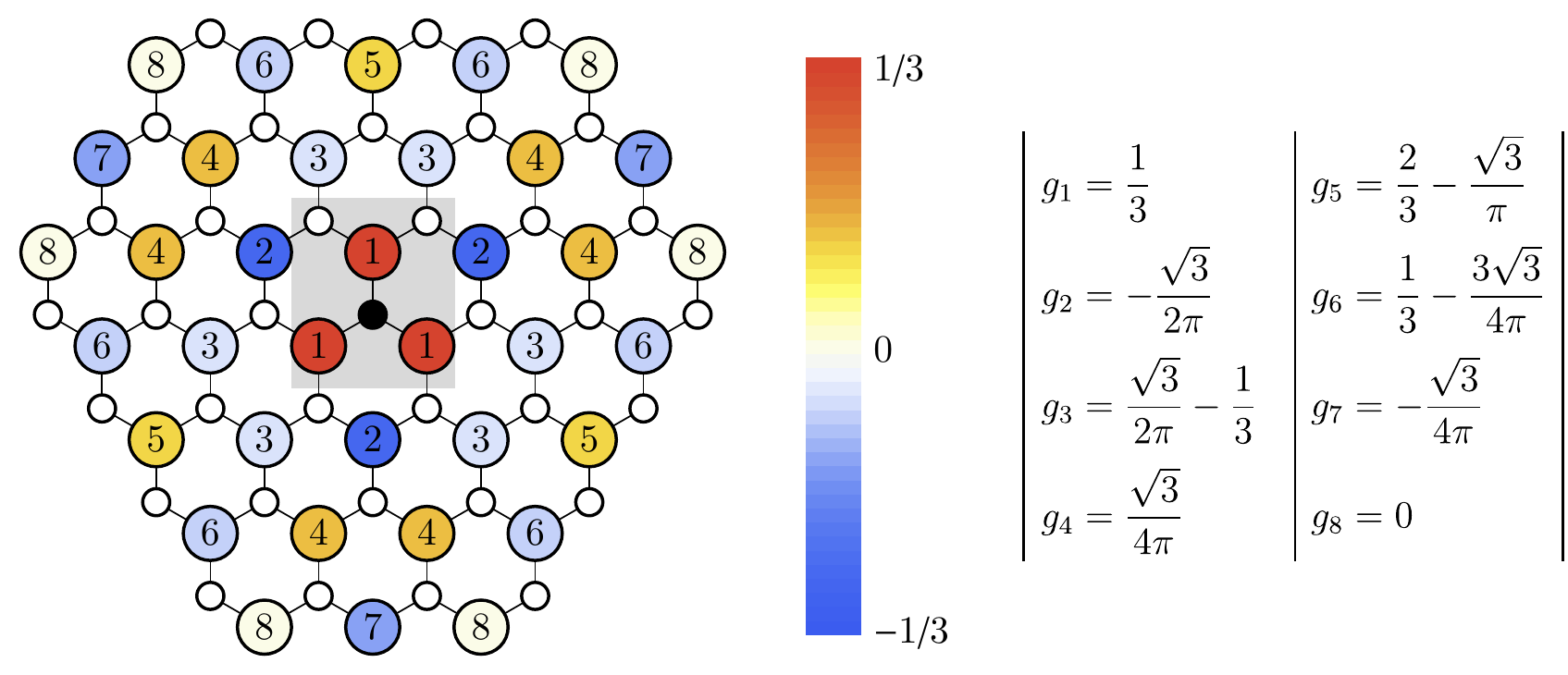}}
\caption{Zero-energy Green's function of the tight-binding Hamiltonian on the honeycomb lattice (up to $1/t$ factor). Small solid circle shows the origin site in the A sublattice. Other sites of the A sublattice are marked with small open circles; the Green function vanishes there. The numbers on the B sublattice sites distinguish different values of the Green's function listed in the right panel and also color-coded. Shaded region shows six neighboring sites perturbed by a structural defect (585 or Stone-Wales). They correspond to the positions listed in Eq.\ (\protect\ref{basis}).}
\label{fig:Green}
\end{figure}

Zero-energy Green's function is simply the inverse of the Hamiltonian operator. It has nonzero matrix elements only between sites from different sublattices. We construct the Green's function in momentum representation and then transform it to real-space as
\begin{equation}
 g\Bigl( n_1 (\bm{\delta}_1 - \bm{\delta}_0) + n_2 (\bm{\delta}_2 - \bm{\delta}_0) + \bm{\delta}_0 \Bigr)
  = t^{-1} \int\limits_0^{2\pi} \frac{dk_1\, dk_2}{(2\pi)^2}\; \frac{e^{i k_1 n_1 + i k_2 n_2}}{1 + e^{i k_1} + e^{i k_2}}.
\end{equation}
The argument of this function is the vector connecting a site in the A sublattice with a site in the B sublattice displaced by $n_{1,2}$ elementary translations in the directions $\bm{\delta}_{1,2} - \bm{\delta}_0$. The first few values of the Green function are displayed in Fig.\ \ref{fig:Green}.

We will consider the Stone-Wales defect and the 585 defect that perturb at most six neighboring sites of the lattice (shaded region in Fig.\ \ref{fig:Green}). We select a basis of these six sites to be
\begin{equation}
 \mathbf{r}_0 + \Bigl\{ 0,\; \bm{\delta}_0 - \bm{\delta}_1,\;  \bm{\delta}_0 - \bm{\delta}_2,\; \bm{\delta}_0,\; \bm{\delta}_1,\; \bm{\delta}_2 \Bigr\}.
 \label{basis}
\end{equation}
The offset position $\mathbf{r}_0$ defines location of the impurity and refers to some site in the A sublattice. Between these six points, the lattice Green's function takes the values
\begin{equation}
 \hat g
  = \begin{pmatrix}
      0 & 0 & 0 & g_1 & g_1 & g_1 \\
      0 & 0 & 0 & g_1 & g_3 & g_2 \\
      0 & 0 & 0 & g_1 & g_2 & g_3 \\
      g_1 & g_1 & g_1 & 0 & 0 & 0 \\
      g_1 & g_3 & g_2 & 0 & 0 & 0 \\
      g_1 & g_2 & g_3 & 0 & 0 & 0
    \end{pmatrix},
 \qquad\qquad
 \begin{aligned}
   g_1 &= \dfrac{1}{3t}, \\
   g_2 &= -\dfrac{\sqrt{3}}{2\pi t}, \\
   g_3 &= \dfrac{1}{t} \biggl( -\dfrac{1}{3} + \dfrac{\sqrt{3}}{2\pi} \biggr).
 \end{aligned}
\end{equation}

\subsection{Dirac Hamiltonian}

Effective description of the low-energy electrons in graphene is provided by the massless Dirac Hamiltonian. In the valley-symmetric form it can be written as
\begin{equation}
 H
  = v \bm{\sigma} \mathbf{p}
  = -i \hbar v \left( \sigma_x \frac{\partial}{\partial x} + \sigma_y \frac{\partial}{\partial y} \right),
 \qquad
 v = \frac{\sqrt{3}\, t a}{2 \hbar} = 10^6\,\text{m/s}.
\end{equation}
Here $\bm{\sigma} = \{\sigma_x, \sigma_y\}$ is a vector of Pauli matrices. The Dirac Hamiltonian acts on four-component wave functions in the space of two sublattices (A and B) and two valleys ($K$ and $K'$). The valleys are defined by the position of the Dirac points in momentum-space, $\mathbf{K} = (4\pi/3a)(\sin\alpha,\; -\cos\alpha)$ and $\mathbf{K}' = -\mathbf{K}$.

The Matsubara Green's function of the Dirac Hamiltonian at an imaginary energy $E = i\epsilon$ is
\begin{equation}
 G(i\epsilon, \mathbf{r})
  = \int \frac{d^2 p}{(2\pi\hbar)^2}\; \frac{i\epsilon + v \bm{\sigma}\mathbf{p}}{\epsilon^2 + v^2 p^2}\; e^{i \mathbf{pr}/\hbar}
  = -\frac{i\epsilon}{2\pi\hbar^2 v^2} \left[
      K_0\left( \frac{\epsilon r}{\hbar v} \right) + \frac{\bm{\sigma}\mathbf{r}}{r} K_1\left( \frac{\epsilon r}{\hbar v} \right)
    \right].
\end{equation}
At short distances $\epsilon r \ll \hbar v$ this function has the asymptotics
\begin{equation}
 G(i\epsilon, \mathbf{r})
  \approx \frac{-i \bm{\sigma}\mathbf{r}}{2\pi\hbar v r^2}
    +\frac{i\epsilon}{2\pi\hbar^2 v^2} \left[
      \log\left( \frac{\epsilon r}{2\hbar v} \right) + \gamma
    \right],
 \label{GDirac}
\end{equation}
where $\gamma \approx 0.577$ is the Euler-Mascheroni constant.

\subsection{Relation between tight-binding and Dirac description}

The four-component wave function $| \Phi(\mathbf{r}) \rangle$ governed by the Dirac Hamiltonian is related to the lattice wave function $\Psi(\mathbf{r})$ in the following way (see Ref.\ \onlinecite{Schelter})
\begin{equation}
 \Psi(\mathbf{r})
  = \langle u(\mathbf{r}) | \Phi(\mathbf{r}) \rangle,
 \qquad
 \langle u(\mathbf{r}) |
  = \sqrt{A} \begin{cases}
      \left( e^{i\alpha/2 + i\mathbf{K r}},\; 0,\; 0,\; e^{-i\alpha/2 - i\mathbf{K r}} \right), & \mathbf{r} \in \text{A}, \\
      \left( 0,\; i e^{-i\alpha/2 + i\mathbf{K r}},\; i e^{i\alpha/2 - i\mathbf{K r}},\; 0 \right), & \mathbf{r} \in \text{B},
    \end{cases}
 \qquad
 A
  = \frac{\sqrt{3}}{2}\, a^2.
\end{equation}
Here $A$ is the area of the unit cell.

For the six basis sites defined in Eq.\ (\ref{basis}) we can form a corresponding $6 \times 4$ matrix composed of the rows $\langle u (\mathbf{r})|$. This matrix is conveniently represented in the factorized form $\bigl\{ \langle u | \bigr\} = W U$ with matrices
\begin{equation}
 W
  = \sqrt{A} \begin{pmatrix}
      1 & 0 & 0 & 1 \\
      e^{2\pi i/3} & 0 & 0 & e^{-2\pi i/3} \\
      e^{-2\pi i/3} & 0 & 0 & e^{2\pi i/3} \\
      0 & 1 & 1 & 0 \\
      0 & e^{-2\pi i/3} & e^{2\pi i/3} & 0 \\
      0 & e^{2\pi i/3} & e^{-2\pi i/3} & 0
    \end{pmatrix},
 \qquad
 U
  = \begin{pmatrix}
      e^{i\alpha/2 + i \mathbf{Kr}_0} & 0 & 0 & 0 \\
      0 & i e^{-i\alpha/2 + i \mathbf{Kr}_0} & 0 & 0 \\
      0 & 0 & i e^{i\alpha/2 - i \mathbf{Kr}_0} & 0 \\
      0 & 0 & 0 & e^{-i\alpha/2 - i \mathbf{Kr}_0}
    \end{pmatrix}.
\end{equation}
The diagonal unitary matrix $U$ encodes dependence on the impurity position $\mathbf{r}_0$ and the lattice orientation $\alpha$.

\subsection{Impurity T matrix}

A single impurity is described by the perturbation $\hat V$ to the tight-binding Hamiltonian. At zero energy, it corresponds to the $T$ matrix defined on the lattice as
\begin{equation}
 \hat t
  = \hat V (1 - \hat g \hat V)^{-1}.
 \label{hatt}
\end{equation}
In the Dirac language, the $T$ matrix becomes
\begin{equation}
 T = U^\dagger T_0 U,
 \qquad
 T_0 = W^\dagger \hat t W.
 \label{T0}
\end{equation}
Here we have also introduced the notation $T_0$ for the $T$ matrix of an impurity placed at $\mathbf{r}_0 = 0$ with the orientation $\alpha = 0$. The general $T$ matrix differs from $T_0$ by the diagonal unitary rotation $U$ only.

At a finite Matsubara energy $E = i\epsilon$, we can express the $T$ matrix using the following identities:
\begin{equation}
 T(i\epsilon) = U^\dagger T_0(i\epsilon) U,
 \qquad
 T_0(i\epsilon)
  = T_0 \bigl[ 1 - \Delta G(i\epsilon) T_0 \bigr]^{-1}.
 \label{TE}
\end{equation}
Here $\Delta G(i\epsilon)$ denotes the difference of two Green's functions of the Dirac Hamiltonian (\ref{GDirac}) taken at coincident points.
\begin{equation}
 \Delta G(i\epsilon)
  = \lim_{\mathbf{r} \to 0} \Bigl[ G(i\epsilon, \mathbf{r}) - G(0, \mathbf{r}) \Bigr]
  \approx \frac{i\epsilon \log(\epsilon/t)}{2\pi\hbar^2 v^2}.
\end{equation}
Here we have used Eq.\ (\ref{GDirac}) and replace $r \mapsto a$ in the argument of the logarithm at short distances.

\subsection{Self energy for weak impurities: Stone-Wales defect}

We first apply the above formalism to graphene with Stone-Wales defects. An isolated impurity is equivalently described by one of the two perturbation operators
\begin{equation}
 \hat V_\text{SW}
  = t \begin{pmatrix}
      0 & -1 & 0 & 0 & 1 & 0 \\
      -1 & 0 & 0 & 1 & 0 & 0 \\
      0 & 0 & 0 & 0 & 0 & 0 \\
      0 & 1 & 0 & 0 & -1 & 0 \\
      1 & 0 & 0 & -1 & 0 & 0 \\
      0 & 0 & 0 & 0 & 0 & 0
    \end{pmatrix}
 \qquad \text{or} \qquad
 \hat V_\text{SW}
  = t \begin{pmatrix}
      0 & 0 & -1 & 0 & 0 & 1 \\
      0 & 0 & 0 & 0 & 0 & 0 \\
      -1 & 0 & 0 & 1 & 0 & 0 \\
      0 & 0 & 1 & 0 & 0 & -1 \\
      0 & 0 & 0 & 0 & 0 & 0 \\
      1 & 0 & 0 & -1 & 0 & 0
    \end{pmatrix}.
 \label{VSW}
\end{equation}
From the perturbation matrix we construct the lattice $T$ matrix $\hat t$ at zero-energy using Eq.\ (\ref{hatt}) and translate it into the Dirac language applying Eq. (\ref{T0}). For the moment we disregard the phase factors $U^\dagger$ and $U$. The resulting $T$ matrix has only two non-zero eigenvalues and can be represented as
\begin{equation}
 T_0
  = W^\dagger \hat V (1 - \hat g \hat V)^{-1} W
  = \frac{3 t A}{2} \begin{pmatrix} 1 & 1 \\ 1 & -1 \\ -1 & -1 \\ -1 & 1 \end{pmatrix}
    \begin{pmatrix} \dfrac{3 \pi}{6 \sqrt{3} - 5 \pi} & 0 \\ 0 & 1 \end{pmatrix}
    \begin{pmatrix} 1 & 1 & -1 & -1 \\ 1 & -1 & -1 & 1 \end{pmatrix}.
\end{equation}
This result is independent of the choice of the $\hat V$ matrix from the two alternatives shown in Eq.\ (\ref{VSW}).

The energy dependence of the $T$ matrix is given by Eq.\ (\ref{TE}). Since we are interested in low energies only, we can expand to the linear order in $\Delta G$ and obtain
\begin{equation}
 T(i\epsilon)
  \approx U^\dagger \bigl[ T_0 + T_0^2 \Delta G(i\epsilon) \bigr] U.
\end{equation}
Now we average over positions and orientations of the impurity. This averaging implies changing the phases contained in $U$ and effectively annihilates all the non-diagonal elements of the $T$ matrix. The result of averaging is
\begin{equation}
 \langle T(i\epsilon) \rangle
  = \frac{3tA}{2} \left( 1 - \frac{3 \pi}{5 \pi - 6 \sqrt{3}} \right)
    + 9 t^2 A^2 \Delta G(i\epsilon) \left[ 1 + \left( \frac{3 \pi}{6 \sqrt{3} - 5 \pi} \right)^2 \right].
\end{equation}
We see that the matrix structure of the $T$ matrix is trivial after averaging.

The corresponding self energy can be written as
\begin{equation}
 \Sigma
  = \frac{n_\text{imp}}{A}\, \langle T(i\epsilon) \rangle
  = n_\text{imp} \Bigl[ \delta + i \alpha \epsilon \log(\epsilon/t) \Bigr].
 \label{SigmaSW}
\end{equation}
Here we have introduced the dimensionless concentration of defects $n_\text{imp}$ assuming the area of a single defect to be equal to the lattice unit cell area $A$. The parameters $\delta$ and $\alpha$ are
\begin{gather}
 \delta
  = \frac{3}{2} \left( 1 - \frac{3 \pi}{5 \pi - 6 \sqrt{3}} \right) t
  = -3.52\,\text{eV}, \\
 \alpha
  = \frac{9 t^2 A}{2\pi\hbar^2 v^2} \left[ 1 + \left( \frac{3 \pi}{5 \pi - 6 \sqrt{3}} \right)^2 \right]
  = \frac{3\sqrt{3}}{\pi} \left[ 1 + \left( \frac{3 \pi}{5 \pi - 6 \sqrt{3}} \right)^2 \right]
  = 6.85.
\end{gather}

Let us note that the values of $\delta$ and $\alpha$ are specific to the Stone-Wales defect while the functional form of the self energy (\ref{SigmaSW}) is universal for any weak (nonresonant) impurities.

\subsection{Self energy for resonant impurities: 585 defect}

The 585 defect can be described by the perturbation matrix
\begin{equation}
 \hat V_{585}
  = t \begin{pmatrix}
      \ast & 0 & 0 & \ast & 1 & 1 \\
      0 & 0 & -1 & 1 & 0 & 0 \\
      0 & -1 & 0 & 1 & 0 & 0 \\
      \ast & 1 & 1 & \ast & 0 & 0 \\
      1 & 0 & 0 & 0 & 0 & -1 \\
      1 & 0 & 0 & 0 & -1 & 0
    \end{pmatrix}.
\end{equation}
The elements marked with $\ast$ are unimportant since the corresponding sites are completely detached from the rest of the lattice.

The 585 defect is special because its zero-energy $T$ matrix diverges. The combination $1 - \hat g \hat V$ has a zero eigenvalue that signifies an emergence of a localized eigenstate. Such impurities are known as resonant \cite{Titov10, Ostrovsky10}. We retain only the corresponding eigenvector in $\hat t$ and represent the matrix as
\begin{equation}
 \hat t
  = \hat V | \psi \rangle M \langle \psi|,
 \qquad
 | \psi \rangle
  = \{0,\, 1,\, 1,\, 0,\, 1,\, 1 \}^T.
\end{equation}
The limit $M \to \infty$ is assumed.

In the Dirac language, the $T$ matrix of a defect placed at $\mathbf{r}_0 = 0$ with orientation $\alpha = 0$ becomes
\begin{equation}
 T_0 = W^\dagger \hat V | \psi \rangle M \langle \psi| W.
\end{equation}
At a finite energy $E = i \epsilon$, we can find the $T$ matrix using Eq.\ (\ref{TE}) and take the limit $M \to \infty$
\begin{equation}
 T(i\epsilon)
  = \lim_{M \to \infty}
    \frac{U^\dagger W^\dagger \hat V | \psi \rangle M \langle \psi | W U}{1 - M \langle \psi | W \Delta G(i\epsilon) W^\dagger \hat V | \psi \rangle}
  = -\frac{U^\dagger W^\dagger \hat V | \psi \rangle \langle \psi | W U}{\langle \psi | W \Delta G(i\epsilon) W^\dagger \hat V | \psi \rangle}
  = -\frac{1}{4\Delta G(i\epsilon)}\; U^\dagger \begin{pmatrix} 1 & 1 & 1 & 1 \\ 1 & 1 & 1 & 1 \\ 1 & 1 & 1 & 1 \\ 1 & 1 & 1 & 1 \end{pmatrix} U.
\end{equation}

Averaging over positions and orientations suppresses all non-diagonal terms. This leads to the following self energy
\begin{equation}
 \Sigma
  = \frac{n_\text{imp}}{A}\, \langle T(i\epsilon) \rangle
  = -\frac{\beta n_\text{imp}}{i\epsilon \log(\epsilon/t)},
 \qquad
 \beta = \frac{\sqrt{3} \pi}{4}\, t^2 = 12.5\,(\text{eV})^2.
\end{equation}

\subsection{Self-consistent T matrix approximation}

Electron wavelength diverges at small energies hence impurities cannot be studied individually. The simplest approach taking into account interference between scattering on different impurities is the self-consistent $T$ matrix approximation. Although this approximation is not quantitatively justified near the Dirac point, it is known to capture all the qualitative features of disordered graphene \cite{Ostrovsky06}.

In the most general setting, we assume simultaneous presence of Stone-Wales and 585 defects and have the following self-consistency equation:
\begin{equation}
 \Sigma
  = n_\text{SW} \alpha (E - \Sigma) \log\bigl[-i(E - \Sigma)/t\bigr] - \frac{\beta n_{585}}{(E - \Sigma) \log\bigl[-i(E - \Sigma)/t\bigr]}.
 \label{SCTMA}
\end{equation}
Here we have restored real energy, $E = i \epsilon$, and neglected the parameter $\delta$ that leads to an unimportant overall energy shift. Equation (\ref{SCTMA}) reduces to Eqs.\ (4) and (5) of the main text when only one type of defect is present.

The above equation determines $\Sigma$ as a function of $E$. We will assume a solution with a negative imaginary part, which corresponds to the retarded self energy. The knowledge of $\Sigma$ allows us to represent the spectral weight
\begin{equation}
 A(E, \mathbf{p})
  = -\frac{1}{\pi} \mathop{\mathrm{Im}} \mathop{\mathrm{tr}} \langle G(E, \mathbf{p}) \rangle
  = -\frac{2}{\pi} \mathop{\mathrm{Im}} \left[ \frac{1}{E - \Sigma - v p} + \frac{1}{E - \Sigma + v p} \right].
\end{equation}
This is Eq.\ (3) of the main text. We see that the spectral weight at a given energy is a sum of two symmetric Lorentz peaks in $p$ centered at $\pm (E -  \mathop{\mathrm{Re}} \Sigma)$ with the width $|\mathop{\mathrm{Im}} \Sigma|$. The total density of states can be obtained as a momentum integral of the spectral weight,
\begin{equation}
 \rho(E)
  = \int \frac{d^2 p}{(2\pi\hbar)^2}\; A(E, \mathbf{p})
  = -\frac{2}{\pi^2 \hbar^2 v^2} \mathop{\mathrm{Im}} \Bigl( (E - \Sigma) \log\bigl[ -i(E - \Sigma)/t \bigr] \Bigr).
\end{equation}
This equation was used to plot $\rho(E)$ in Fig.\ 4e of the main text.

\end{document}